\documentclass[fleqn,usenatbib]{mnras}

\usepackage{newtxtext,newtxmath}

\usepackage[T1]{fontenc}

\DeclareRobustCommand{\VAN}[3]{#2}
\let\VANthebibliography\thebibliography
\def\thebibliography{\DeclareRobustCommand{\VAN}[3]{##3}\VANthebibliography}

\usepackage{graphicx}	
\usepackage{amsmath}	
\usepackage{hyperref}
\usepackage{float}
\usepackage{xcolor}
\usepackage{ulem} 
\usepackage{caption}

\title[Cluster mass distribution $\&$ CMB lensing]{A CMB lensing analysis of the extended mass distribution of clusters}

\author[Facundo Toscano et al.]{
Facundo Toscano$^{1}$\thanks{\textbf{E-mail:} facundo.toscano@mi.unc.edu.ar},
Heliana Luparello$^{1}$,
Elizabeth Johana Gonzalez,$^{2,1,3}$\thanks{also at Port d'Informació Científica (PIC), Campus UAB, C. Albareda s/n, 08193 Barcelona (Barcelona), Spain.} and Diego Garcia Lambas$^{1,3}$
\\
$^{1}$Instituto de Astronomía Teórica y Experimental (IATE-CONICET), Laprida 854, X5000BGR, Córdoba, Argentina. \\
$^{2}$Institut de Física d’Altes Energies (IFAE), The Barcelona Institute of Science and Technology, Campus UAB, 08193 Barcelona, Spain\\
$^{3}$Observatorio Astronómico de Córdoba, Universidad Nacional de Córdoba (OAC-UNC), Laprida 854, X5000BGR, Córdoba, Argentina
}

\date{Accepted 2023 October 05. Received 2023 September 13; in original form 2023 July 11}
\pubyear{2023}

\begin{document}
\label{firstpage}
\pagerange{\pageref{firstpage}--\pageref{lastpage}}
\maketitle

\begin{abstract}
The aim of this work is to study the anisotropic weak lensing signal associated with the mass distribution of massive clusters of galaxies using the Cosmic Microwave Background (CMB) data. For this purpose, we stack patches of the Planck Collaboration 2018 CMB lensing convergence map centered on SDSS DR8 redMaPPer clusters within the redshift range $[0.4, 0.5]$.
We obtain mean radial profiles of the convergence parameter $\kappa$ finding strong signals at scales as large as $40$ Mpc$h^{-1}$. By orienting the clusters along their major axis defined through the galaxy member distribution, we find a significant difference between the parallel and perpendicular oriented convergence profiles. The amplitude of the profile along the parallel direction is about $50\%$ larger than that along the perpendicular direction, indicating that the clusters are well aligned with the surrounding mass distribution. From a model with an anisotropic surface mass density, we obtain a suitable agreement for both mass and ellipticities of clusters compared to results derived from weak lensing shear estimates, finding strong evidence of the correlation between the galaxy cluster member distribution and the large--scale mass distribution.
\end{abstract}

\begin{keywords}
Cosmic background radiation -- large-scale structure of Universe -- gravitational lensing: weak -- galaxies: clusters -- methods: data analysis
\end{keywords}



\section{Introduction}

Cluster masses have been widely used to restrict cosmological parameters \citep{2012AstL...38..347B,2014MNRAS.439.2485C}. In fact, the abundance of clusters above a given mass threshold provides a powerful constrain in $\Omega_M$ and $\sigma_8$ parameters \citep[e.g.,][]{2022A&A...665A.100L, 2022arXiv221009530A}. \\
Among the different methods to infer cluster masses, weak lensing techniques give reliable estimations since they are less affected by the dynamical state of the clusters \citep{2019MNRAS.482.1352M,2022MNRAS.512.4785M}. Besides, unlike other techniques such as those based on X-ray measurements, the weak lensing method has the advantage of being unaffected by astrophysical sources in the clusters which could bias estimates. \\
In numerical simulations, clusters show triaxial shapes \citep{2022MNRAS.513.2178H}, elongated mainly due to the anisotropic large-scale structure and the related accretion process.  For this reason, to accurately reproduce the measured density distribution, a two halo anisotropic term must be included \citep{2022MNRAS.517.4827G}. On the observational side, however, cluster shapes can be only obtained in projection through different tracers such as the member galaxy distribution, intracluster light, X-rays hot emitting gas as well as weak lensing. 
Previous weak lensing studies based on shear measurements \citep{2016MNRAS.457.4135C,Uitert2017,2018MNRAS.475.2421S,2021MNRAS.501.5239G}
have added evidence for cluster halos with an average ellipticity parameter $\epsilon = \frac{1-q}{1+q} \sim 0.2$, where $q$ is the semi-axes relation, $q \leq 1$. These studies are mainly focused on the analysis of the shape of the cluster halo host, neglecting the contribution of the extended mass distribution.  On the other hand, related with the surrounding mass distribution in galaxy clusters, \citet{2013ApJ...764...58G,2018MNRAS.478.5366F,2021MNRAS.507.5758S} have analysed the expected lensing signal at intermediate scales without fitting the expected elongation at these regions. In this sense, the CMB lensing signal can be useful to explore the surrounding mass distribution given its large angular extension.\\
In this paper, we study the convergence map derived from the Planck 2018 lensing data \citep{Planck:2020b} around massive clusters of the SDSS DR8 redMaPPer catalogue \citep{redmapper}. This  catalogue provides an homogeneously selected sample of massive clusters $M \gtrsim 10^{14} M_{\odot}$, suitable for our analysis. We explore the anisotropies imprinted in the CMB convergence map due to the elongation of the extended surface mass distribution, by orientating the maps taking into account the positions of the cluster galaxy members. In this way, we study the anisotropies of the mass distribution beyond the virial region, in close relation to the large--scale structure. \cite{geach2017} also study the redMaPPer clusters catalogue using the Planck 2018 lensing data. Unlike the present work, the authors analyse smaller CMB reprojected regions of $1^{\circ}$ around the position of each cluster onto a tangential sky projection with a high resolution scale of 256 pixels per degree. For this reason, it is observed only a significant signal in the cluster center neighborhood, up to 10 arcmin or around $\sim 3$ Mpc$h^{-1}$. Here, we focus on a large--scale study with the aim of analysing extended anisotropic mass distribution around clusters. \\
The paper is organized as follows: in Section \ref{Data} we detail the data of the CMB lensing map and the cluster catalogue. In Section \ref{analysis} we analyse the mean convergence profiles. In Section \ref{model_sec} we describe an anisotropic model that allows us to trace the extended mass distribution and also explore mean masses and ellipticities of the systems. Finally, in Section \ref{Conclusions} we present a brief summary and discussion of our studies.

\section{Data} \label{Data}

\subsection{CMB Convergence Map} \label{CMB}

\begin{figure} 
	\includegraphics[width=\columnwidth]{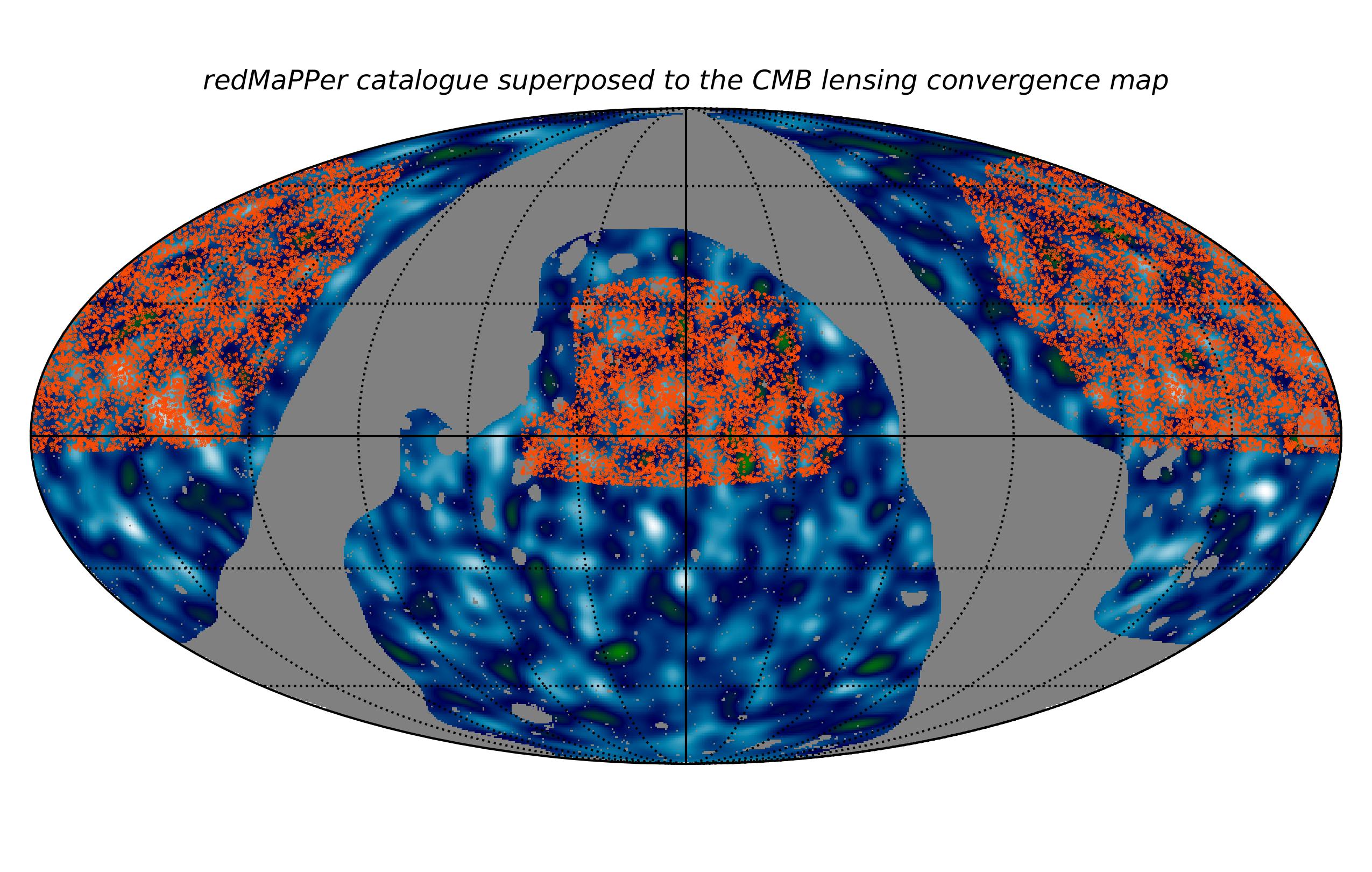}
    \caption{redMaPPer cluster positions (red points)
    superposed to the CMB lensing convergence map in equatorial mollview projection. The lensing map has a FWHM Gaussian smoothing of $5^{\circ}$ for a better visualization. The mask provided by PR3 associated to galactic and extragalactic sources is shown in grey. }
    \label{fig:CMB_cluster}
\end{figure}

In the thin lens approximation \citep[see for instance][and references therein]{2022iglp.book.....M}, the Laplacian of the effective lens potential in the lens plane takes the form:
\begin{equation} \label{Phi}
    \hat{\Psi}(\vec{\theta}) = {{2D_{LS}} \over{c^2D_L D_S}} \int{\Phi (D_L,\vec{\theta},z)}dz
\end{equation}
where $\Phi$ is the tridimensional Newtonian potential defined at a given position in the lens plane, $\vec{\theta}$, and at redshift $z$. $D_L$, $D_S$, $D_{LS}$ are the angular diameter distances from the observer to the lens, from the observer to the source and from the lens to the source, respectively, involved in the lens system. 
The adimensional convergence parameter $\kappa$ is proportional to the effective lens potential by:
\begin{equation} \label{v4}
    \begin{split}
        \bigtriangleup_x \Psi ({\vec{x}}) = 2 \kappa (\vec{x}) 
    \end{split}
\end{equation}
which is related to the surface mass density through
\begin{equation}
        \kappa({\vec{x}}) = {{\Sigma(\vec{x})}\over {\Sigma_{CR}}} \
\end{equation}
where $\Sigma_{CR}$ is the critical surface density:
\begin{gather*}
    \Sigma_{CR} = {{c^2 D_S}\over{4\pi G D_L D_{LS}}}
\end{gather*}
The observed CMB photons arises from the last-scattering surface, and along their path, each wavelength are affected by both the expansion of the Universe as well as by the mass inhomogeneities \citep{1987A&A...184....1B}. Once the structure in the Universe has grown to form, among other systems, clusters of galaxies, each photon crosses several lenses generated by these systems, resulting in a correlation between certain multipoles $l$ that would otherwise have been independent of each other and therefore secondary non Gaussian anisotropies \citep{PhysRevD.58.023003} are introduced. Given this, for the CMB photons the effective lens potential represents an integrated measurement of the mass distribution from the observer to the last scattering surface:
\begin{equation} \label{psi}
    \psi(\mathbf{\hat{n}}) = -2\int_0^{\chi_*}{d\chi \left( {{\chi_*-\chi}\over{\chi_*\chi}} \right) \Psi(\chi ,\mathbf{\hat{n}})}
\end{equation}
where $\chi_*$ is the comoving distance up to the last scattering surface and $\Psi(\chi ,\mathbf{\hat{n}})$ is the Weyl potential, in comoving coordinates. \\
Following the formalism proposed in \cite{2016A&A...594A..15P}, which involves Equations \ref{v4} and \ref{psi}, the Planck Collaboration provides the $k_{LM}$ coefficients:
\begin{gather} \label{k_lm}
    \kappa_{LM} = {{L(L+1)}\over{2}}\psi_{LM}
\end{gather}
which allows to reconstruct the CMB convergence map.
\\
In our analysis, we use the data products released in 2018 by the Planck Collaboration \citep{Planck:2020b}, hereafter PR3. We reconstruct the map from the spherical harmonics coefficients $k_{LM}$ provided by the PR3 release \footnote{\url{https://pla.esac.esa.int/\#home}}, using all the $L's$ available ($L_{Max}=4096$) with a FWHM Gaussian kernel of $1^{\circ}$, which is consistent with the range of $L's$ used in CMB power spectrum analyses \citep{2019PhRvD.100l3509M, 2020A&A...641A...8P}. 
These coefficients are derived from SMICA DX12 CMB maps (from temperature only after mean field subtraction, TT)\footnote{\url{https://wiki.cosmos.esa.int/planck-legacy-archive/index.php/Lensing}}.
For an appropriate analysis, we also use the confidence mask provided by PR3 release which eliminates the most relevant astrophysical effects of galactic and extragalactic objects imprinted in the temperature CMB maps.\\
For the reconstruction and visualization, we use the \texttt{HEALPix} \footnote{\url{https://healpix.sourceforge.io/}} software. It allows us to transform the $k_{LM}$ coefficients to a map with a resolution given by $N_{side}=2048$. The pixel area has a constant value of $A_P= 1.062 \times 10^{4}$ arcsec$^2$. \\
In order to further check the consistency of our methods and results, we also use the PR3 minimum variance (MV) $k_{LM}$ and the TT and MV $k_{LM}$ reconstructed coefficients \citep{carron2022} corresponding to the NPIPE processing pipeline \citep{PR4}(hereafter PR4).\\
In Figure \ref{fig:CMB_cluster}, we present the CMB lensing map reconstruction, from PR3 TT $k_{LM}$ coefficients, with a FWHM Gaussian kernel smoothing of $5^{\circ}$ together with the distribution of redMaPPer clusters, which will be presented in the next section. 

\subsection{Cluster Samples} \label{Clusters} 

\begin{figure} 
	\includegraphics[width=\columnwidth]{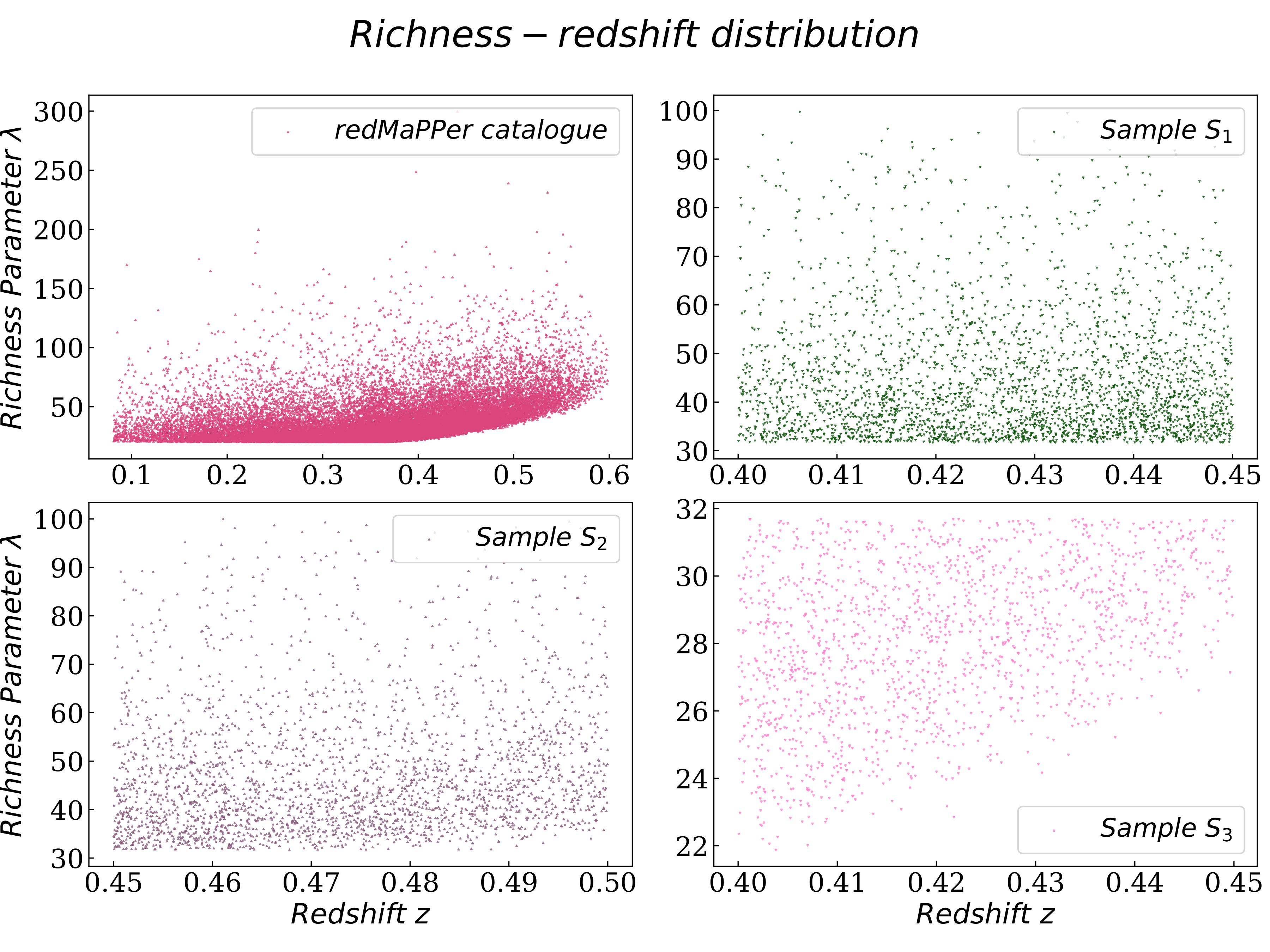}
    \caption{Richness-redshift ($\lambda - z$) distribution. \textit{Upper Left Panel:} redMaPPer catalogue. It can be seen an excess of high values of $\lambda$ at higher $z$. \textit{Upper Right Panel:} Sample $S_1$. Richness is distributed nearly homogeneous across the redshift interval. \textit{Lower Left Panel:} Sample $S_2$. Here, richness is also distributed nearly homogeneous across the redshift interval. \textit{Lower Right Panel:} Sample $S_3$. The observed  excess of high values of $\lambda$ at higher $z$ can induce a bias in the convergence analysis.}
    \label{fig:z_vs_l}
\end{figure}

We analyse samples derived from the redMaPPer cluster catalogue \cite{redmapper} which are identified through a red-sequence algorithm \citep{2014ApJ...785..104R} applied to galaxy over-densities inferred from a percolation procedure on the Sloan Digital Sky Survey Data Release 8 \citep{2000AJ....120.1579Y,sdss-dr8}. This catalogue comprises $26111$ galaxy systems with an angular coverage of $10401$ square degrees in the redshift range $[0.08, 0.6]$.  The catalogue provides a richness parameter, $\lambda$, which is a suitable proxy of cluster mass \citep[e.g.,][]{2017MNRAS.466.3103S,2018MNRAS.474.1361P}. This parameter, ranging between 20 and 300, is defined as the sum of the galaxy cluster membership probabilities over all galaxies within a scale-radius $R_{\lambda}$:
\begin{equation}
    \lambda = \sum p_{mem}\theta_L\theta_R
\end{equation}
where $\theta_L$ and $\theta_R$ are weights associated to a fixed threshold in luminosity and radius.\\
In our analysis we restrict both redshift $z$ and richness parameter $\lambda$ ranges to define several samples which are presented in Table \ref{tab:samples_table} together with their main characteristics. The mean mass $\langle M \rangle$ is calculated according to the relation given by \cite{2017MNRAS.466.3103S}. Also, we consider clusters within the CMB lensing mask region. \\
With the purpose of defining homogeneous samples not strongly affected by redshift nor richness bias, we first consider the redshift range $0.4 \leq z < 0.45$ and select two richness ranges: $21 \leq \lambda \leq 31.7 $ for sample $S_3$ and $31.7 < \lambda \leq 100$ for sample $S_1$. We adopt the threshold $\lambda = 31.7$ which corresponds to the $3000$ richer clusters of sample $S_1$. Additionally, we impose an upper limit  $\lambda \leq 100$ to avoid spuriously high $\lambda$ values probably associated to biases in the percolation procedure. Then, we consider the same richness interval ($31.7 < \lambda \leq 100$) with a different redshift range: $0.45 \leq z \leq 0.5$ for sample $S_2$. We avoid the use of clusters with redshift lower than $0.4$ which are not well suitable for this analysis due to their lower lensing signal.\\
In the upper left panel of Figure \ref{fig:z_vs_l}, we show the richness-redshift ($\lambda - z$) distribution for the catalogue which is complete in richness up to $z \sim 0.4$. In the upper right panel, we show the sample $S_1$. As it can be seen in this figure, this restricted area provides a quasi-homogeneous distribution of richness parameter $\lambda$ across the redshift interval. We present sample $S_2$ in the lower left panel, where we can see that this region also provides a nearly homogeneous distribution across redshift, similarly to sample $S_1$. In the lower right panel, we display sample $S_3$. This sample, contrary to $S_1$ and $S_2$, is not complete in richness.\\
\begin{table*}
\centering
\caption{Main characteristics of cluster samples. The mean mass $\langle M \rangle$ is calculated following the procedure in \citet{2017MNRAS.466.3103S}.}
\label{tab:samples_table}
\begin{tabular}{lcccccr} 
		\hline
		Sample & Number of Clusters & Redshift $z$ range & $\langle z \rangle$ & Richness parameter $\lambda$ range & $\langle \lambda \rangle$ & $\langle M \rangle$  $[M_{\odot}]$ \\
		\hline
		$S_1$ & $3000$ & $[0.4, 0.45]$ & $0.427$ & $[31.7,  100]$ & $44.914$ & $10^{14.304}$ \\
		$S_2$ & $2774$ & $[0.45, 0.5]$ & $0.475$ & $[31.7, 100]$ & $47.818$ & $10^{14.374}$ \\
            $S_3$ & $1574$ & $[0.4, 0.45]$ & $0.420$ & $[21, 31.7]$ & $27.797$ & $10^{14.027}$ \\
		\hline
\end{tabular}
\end{table*}
Some main features can be deduced from Table \ref{tab:samples_table} and Figure \ref{fig:z_vs_l}. It can be seen that $S_1$ and $S_2$ samples have similar values of $\langle M \rangle$ and $\lambda$. For this reason, both samples are well suited for our  analysis. The inhomogeneous distribution of redshift and richness parameters in sample $S_3$ makes this sample unsuitable for a study with well controlled parameters.

\section{Analysis} \label{analysis}
In this section we briefly detail the procedures used to derive mean convergence radial profiles centered on clusters. We also compute oriented profiles according to the parallel and perpendicular directions as defined by the galaxy member distribution.
In order to avoid complications related to both SZ imprint on the CMB and great uncertainties due to  low statistics fluctuations, we have excluded in the analysis the virialized region around the clusters. Therefore, we consider scales beyond $3$ Mpc$h^{-1}$. 

\subsection{Mean radial $\kappa$ profiles} \label{Stacking}

\begin{figure} 
	\includegraphics[width=\columnwidth]{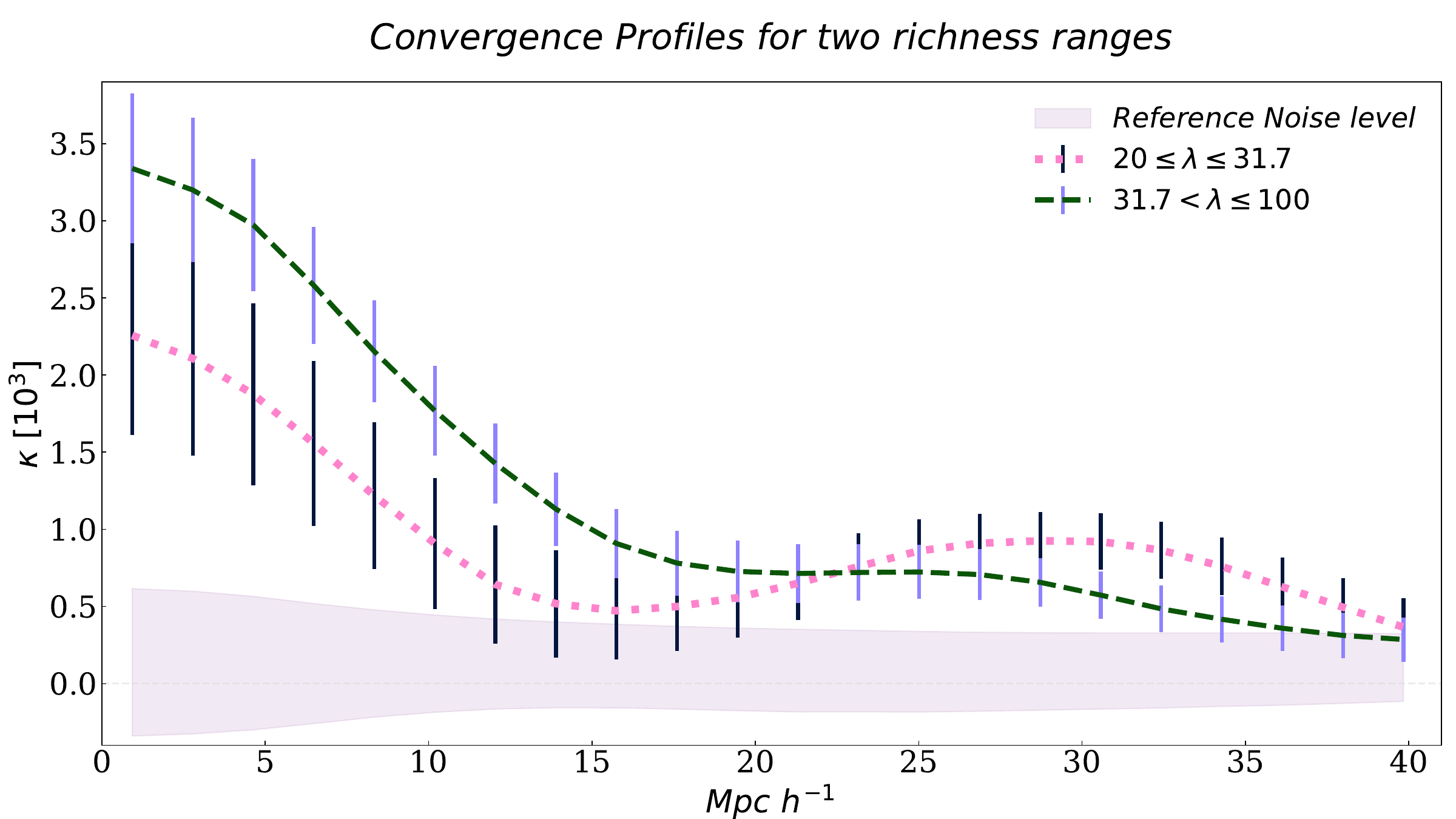}
    \caption{Convergence Profiles for two richness ranges with its uncertainties. In green, the profile corresponding to sample $S_1$. In pink, the profile corresponding to sample $S_3$. It can be seen that the profiles differ significantly up to $15$ Mpc$h^{-1}$, where the lower signal of sample $S_3$ implies that the inclusion of low richness clusters produces a less significant convergence profile. A reference noise level from sample $S_1$ for the CMB convergence map is shown in grey shaded region.}
    \label{fig:Main_low_rich}
\end{figure}

\begin{figure} 
	\includegraphics[width=\columnwidth]{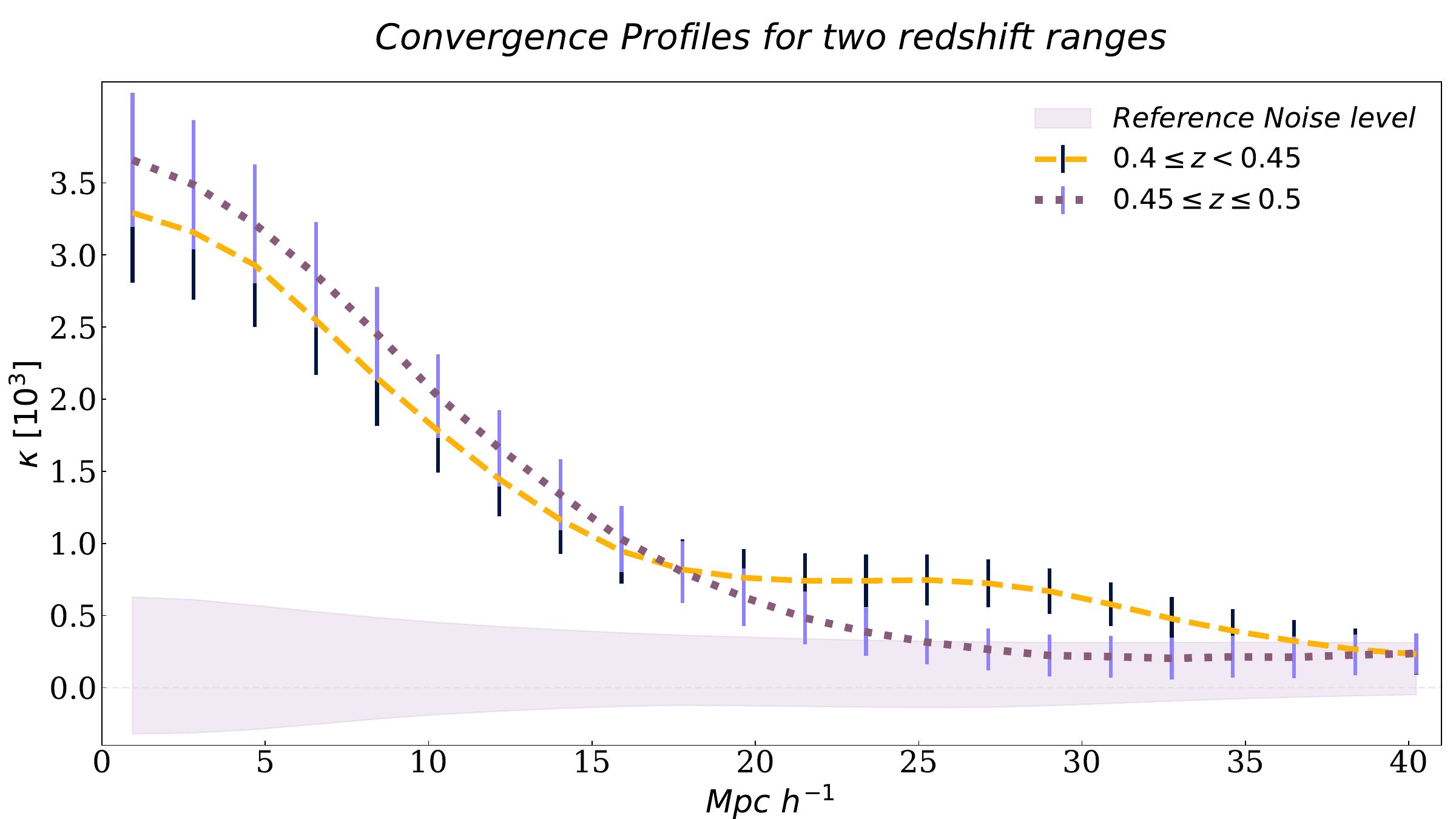}
    \caption{Convergence profiles for two redshift ranges with its uncertainties. In orange, the profile corresponding to sample $S_1$. In violet, the profile corresponding to sample $S_2$. It can be seen that both profiles have a reasonable agreement up to $25$ Mpc$h^{-1}$. A reference noise level from sample $S_2$ for the CMB convergence map is shown in grey shaded region.}
    \label{fig:Two_redshift}
\end{figure}

\begin{figure} 
	\includegraphics[width=\columnwidth]{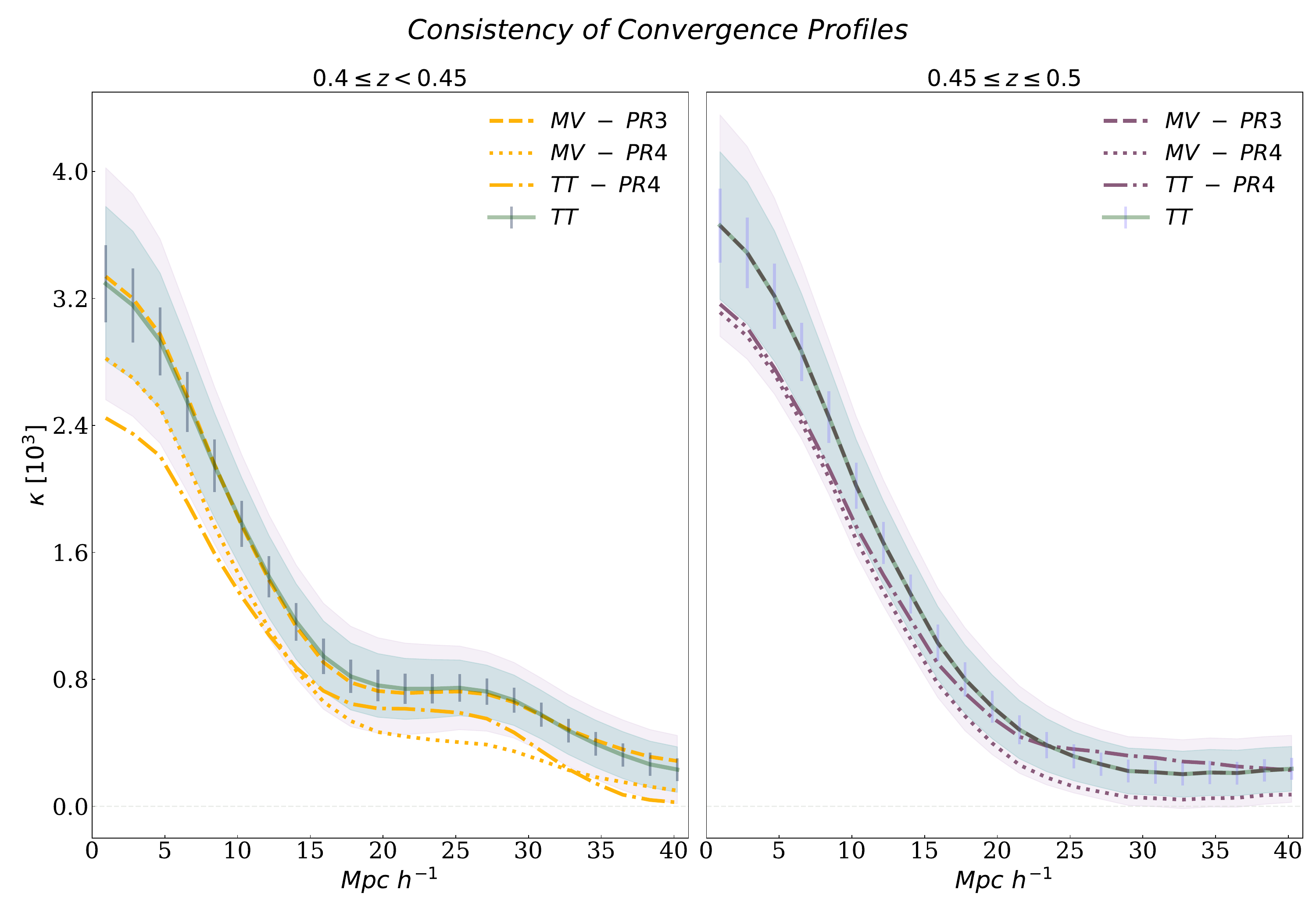}
    \caption{Consistency of the Convergence profiles for samples $S_1$ (yellow curves) and $S_2$ (purple curves). The green solid line corresponds to the PR3 TT map. The dashed, dotted and dash-dotted lines corresponds to PR3 MV, PR4 MV and PR4 TT maps, respectively. $1\sigma$ (errorbar), $2\sigma$ (green shaded) and $3\sigma$ (pink shaded) variances corresponding to PR3 TT map are shown.}
    \label{fig:consistency}
\end{figure}

For our study of the convergence parameter $\kappa$ around clusters, we compute mean radial profiles centered on the clusters. Given that the determination of the $\kappa$ profile for a single cluster has low signal-to-noise ratio, to obtain a smooth $\kappa$ mean radial profile it is necessary to stack several convergence map patches centered on the clusters of a given sample. This is accomplished by superposing map patches around each cluster, where their radii is scaled according to the cluster redshift as a function of the projected distance $r_p$ (in units of Mpc$h^{-1}$) to the cluster center. Over the stacked map we compute the mean $\kappa$ profile for $8$ arcmin radial bins, also scaled according to the cluster redshift, obtaining profiles as a function of distance for each sample.\\
We calculate profiles uncertainties through the bootstrap resampling technique, with $300$ re-samples for each sample considered. Given that for small separations the number of pixels is low, we expect the statistical uncertainty in this region to be larger. Also, we calculate a CMB map noise level. To accomplish this, we compute $300$ new samples with the same number of objects as the sample considered albeit centered in random positions within the redMaPPer region. This map noise level provides a suitable estimate of both the average intrinsic fluctuations in $\kappa$ as well as the distances at which the results are statistically significant. \\
As a test of our procedures and the reliability of the expected results, we have analysed the stability of the method in the three samples previously described (see Table \ref{tab:samples_table}). 
Firstly, we study the dependence of the mean $\kappa$ profiles with respect to the cluster richness $\lambda$. The results are shown in Figure \ref{fig:Main_low_rich}, where we separate the richness in two different ranges: $21 \leq \lambda \leq 31.7$ (sample $S_3$) and $31.7<\lambda \leq 100$ (sample $S_1$). These intervals correspond to $\sim 35 \%$ and $\sim 65 \%$ of the clusters, respectively. It can be seen the significantly larger convergence amplitude for the higher richness clusters up to $15$ Mpc$h^{-1}$, as expected due to the tight correlation between cluster mass and richness.
Given its larger signal, in what follows we will consider this cluster richness threshold (samples $S_1$ and $S_2$, see Table \ref{tab:samples_table} for details). 
In Figure \ref{fig:Two_redshift} we show mean $\kappa$ profiles for two redshift ranges: $0.4 \leq z < 0.45$ (sample $S_1$) and $0.45 \leq z \leq 0.5$ (sample $S_2$). As seen in this figure, the mean radial profiles obtained for the different redshifts bins are mutually consistent, in particular in the region up to $25$ Mpc$h^{-1}$, as compared to the samples $S_1$ and $S_3$ which differ in richness. Also, it can be seen that the profile obtained for sample $S_2$ has a  more gentle decay than $S_1$. We argue that these two behaviors are related to the different structures surrounding the cluster samples, influencing the radial profiles at large distances. \\
In order to test the consistency of our results, we explore the convergence profiles of samples $S_1$ and $S_2$ in three different CMB lensing reconstructed maps using the PR3 MV, PR4 MV and PR4 TT $k_{LM}$ coefficients. In Figure \ref{fig:consistency} it can be seen these profiles compared to the PR3 TT profile and the corresponding statistical confidence regions. It can be noticed that in the PR3 release there is no difference between TT and MV reconstructed maps profiles. In addition, it can be inferred that there is not a statistically significant difference between the PR4 TT and MV profiles, and the PR3 TT profile.  

\subsection{Oriented $\kappa$ profiles} \label{Orientation}

\begin{figure}
        \centering
	\includegraphics[width=\columnwidth]{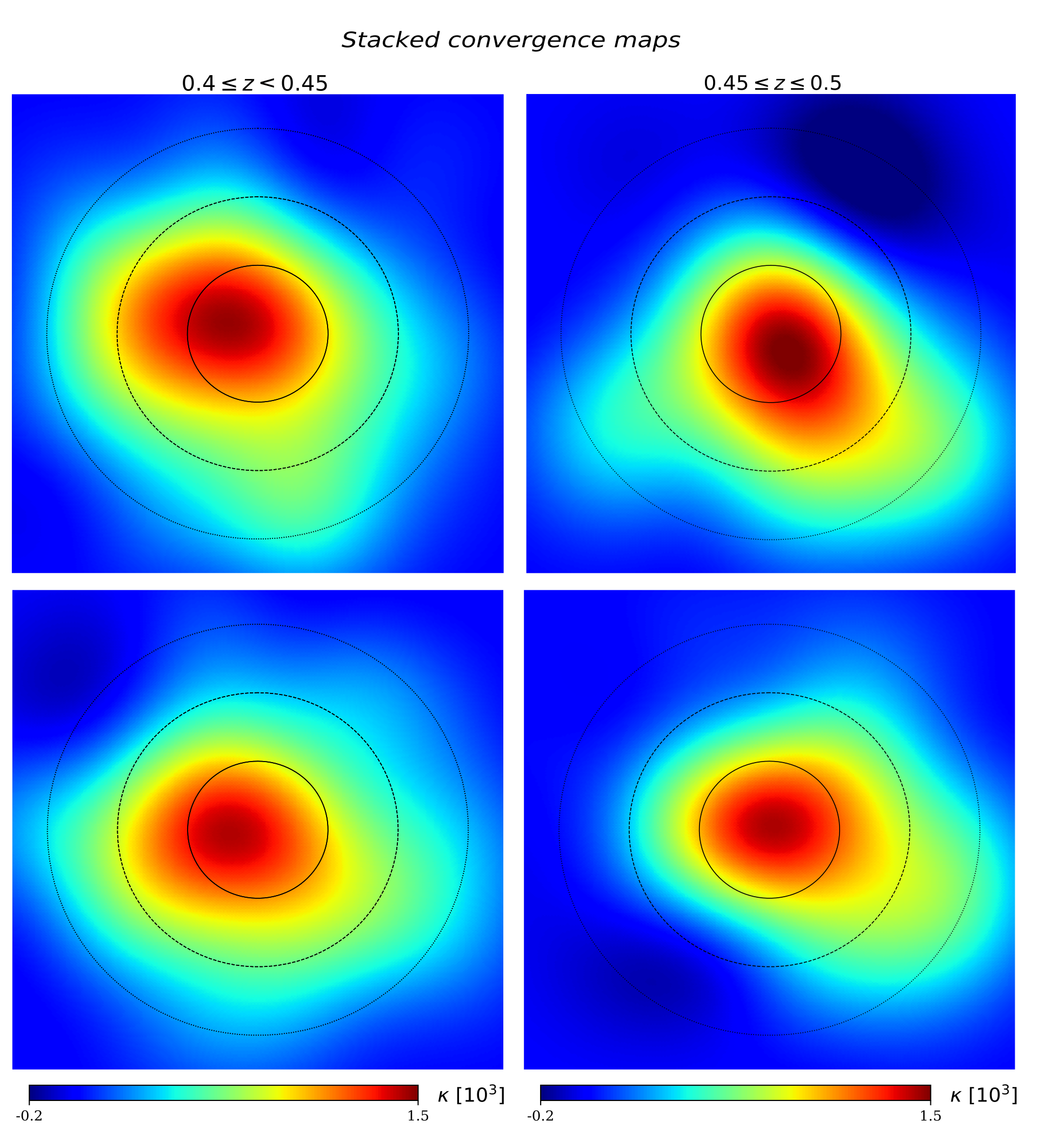}
    \caption{Stacked convergence maps for samples $S_1$ (left panels) and $S_2$ (right panels) with a smoothing corresponding to a FWHM Gaussian kernel of $2^{\circ}$, for a better visualization. The upper panels show the stacked maps without alignment. The lower panels show the stacked aligned maps. The black circles correspond to an angular scale of $1^{\circ}$ (solid), $2^{\circ}$ (dashed) and $3^{\circ}$ (dotted). It can be seen in the aligned stacked maps that the convergence values are greater in the parallel orientation than in the perpendicular ones.}
    \label{fig:stacked}
\end{figure}

\begin{figure*} 
	\includegraphics[width=0.8\textwidth]{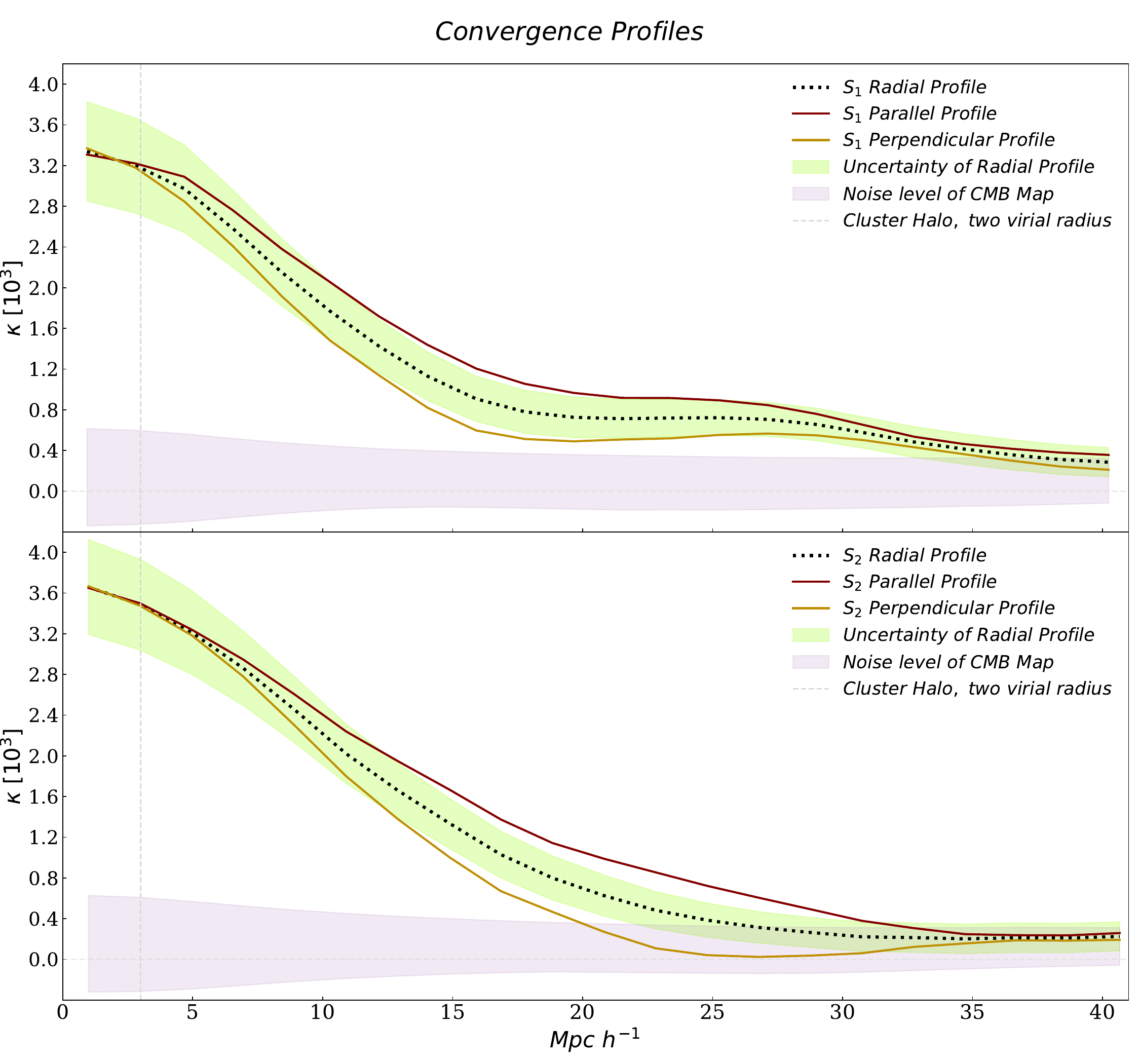}
    \caption{Convergence lensing parameter $\kappa$ profile for sample $S_1$ (upper panel) and $S_2$ (lower panel) as a function of the projected distance to the cluster centres (black dotted line). Brown and yellow curves correspond to parallel and perpendicular directions as defined previously in the text, respectively. The green shaded region around the mean radial profile correspond to the radial uncertainty calculate through a bootstrap resampling technique with $300$ re-samples from the considered sample. The horizontal grey shaded region is the map noise level calculated through $300$ random profiles. The vertical dotted line represent two mean virial radius ($\sim 3$ Mpc$h^{-1}$) where a significant SZ effect is expected.}
    \label{fig:K_total}
\end{figure*}

In order to derive $\kappa$ profiles oriented along the cluster projected major axis, we use the same procedure as in \citet{2018MNRAS.475.2421S} and \cite{2021MNRAS.501.5239G} to calculate the position angles of our cluster samples. We consider an uniform weight $w_k = 1$ and all the galaxy members. Then we stack convergence map patches rotated according to these angles. The stacked aligned $\kappa$ patches are shown in Figure \ref{fig:stacked} for samples $S_1$ and $S_2$, respectively. For comparison, we also plot the non-aligned stacked patches. It can be seen that the $\kappa$ distribution is preferentially oriented along the cluster major axis direction $x$ after the clusters alignment. We notice the miscentering in sample $S_1$, as well as a larger departure from a regular shape in $S_2$. We argue that these two effects are mainly due to statistical fluctuations arising from both smoothing and large--scale irregularities in the mass distribution. This preferred alignment effect extends to several cluster radii up to 40 Mpc$h^{-1}$, which corresponds to $\sim 3^{\circ}$. In both cases, it can be seen that in the cluster nearby regions, $<1^{\circ}$, the convergence is nearly isotropic unlike the outer regions where the convergence tends to show strongly elongated shapes. We also notice a low statistical significance beyond $2^{\circ}$ ($\sim 25$ Mpc$h^{-1}$), as expected from the noise level of the convergence profiles in Figures \ref{fig:Main_low_rich} and \ref{fig:Two_redshift}.\\
In order to highlight the alignment between the cluster orientation and the underlying mass distribution we have computed the convergence profile for pixels within $ 45^{\circ}$ from the cluster major axis direction, hereafter parallel direction. Similarly, we associate the perpendicular direction to the convergence profile for pixels with relative angles $> 45^{\circ}$.   
The resulting parallel and perpendicular convergence $\kappa$ profiles for samples $S_1$ and $S_2$ are shown in Figure \ref{fig:K_total}. For reference, we also plot the total (parallel and perpendicular) radial profile shown and analyzed in the previous section (see Figure \ref{fig:Two_redshift}). The noise level corresponds to the radial profile. 
The vertical dotted line correspond to the virialized clusters region ($\sim 3$ Mpc$h^{-1}$) with a large $\kappa$ uncertainty determination due to the presence of SZ imprint. \\
It can be seen that for both samples the parallel profile has a significantly larger signal in a wide range of scales (from $\sim 5$ Mpc$h^{-1}$ to $40$ Mpc$h^{-1}$) than the perpendicular ones. This behavior is associated to the strong alignment between the inner non-linear virialized region (where the galaxy members use to calculate the position angles are located) and the large--scale mass distribution as provided by the convergence $\kappa$ map. It is remarkable the fact that the alignment relates the distribution of galaxies, within $\sim 1-2$ Mpc$h^{-1}$, and the mass density fluctuations at scales of at least $40$ Mpc$h^{-1}$. This is particularly significant for the higher redshift bin (sample $S_2$). 

\section{Anisotropic large-scale mass distribution model} \label{model_sec}

\begin{figure}
	\includegraphics[width=\columnwidth]{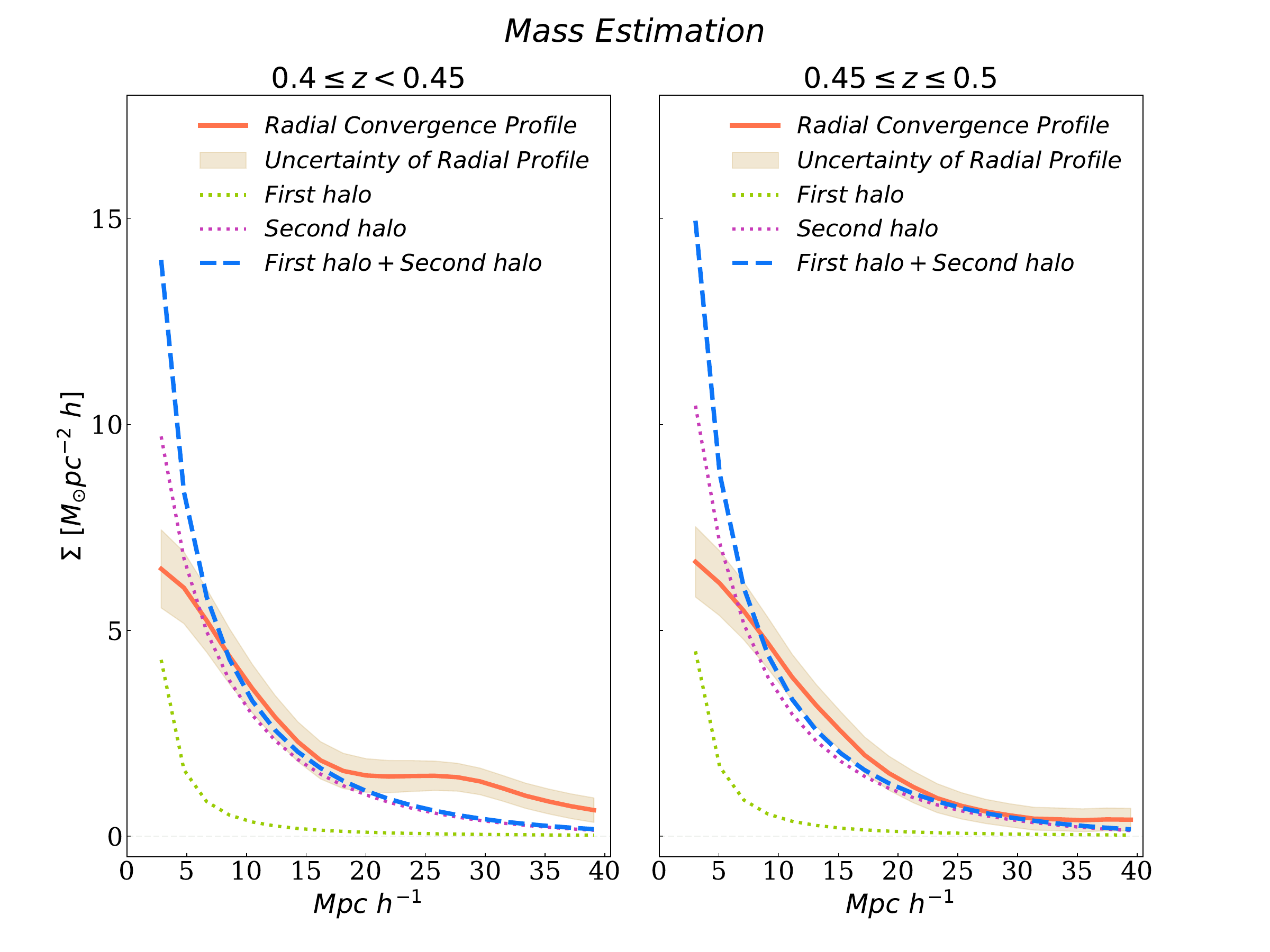}
    \caption{Density monopolar model compared to mean radial convergence profile for samples $S_1$ (Left Panel) and $S_2$ (Right Panel). In dashed blue, it can be seen the sum of the first halo term (dotted green) and second halo term (dotted pink) of the anisotropic density model calculated through $\langle M \rangle$ of samples $S_1$ and $S_2$. In orange, the mean radial convergence profiles with its uncertainty. It can be seen the good agreement of the model with respect to the measurements. The results on lower scales than $5$ Mpc$h^{-1}$ have great statistical uncertainties and larger contamination of SZ effect.}
    \label{fig:z_vs_l1}
\end{figure}

\begin{figure} 
	\includegraphics[width=\columnwidth]{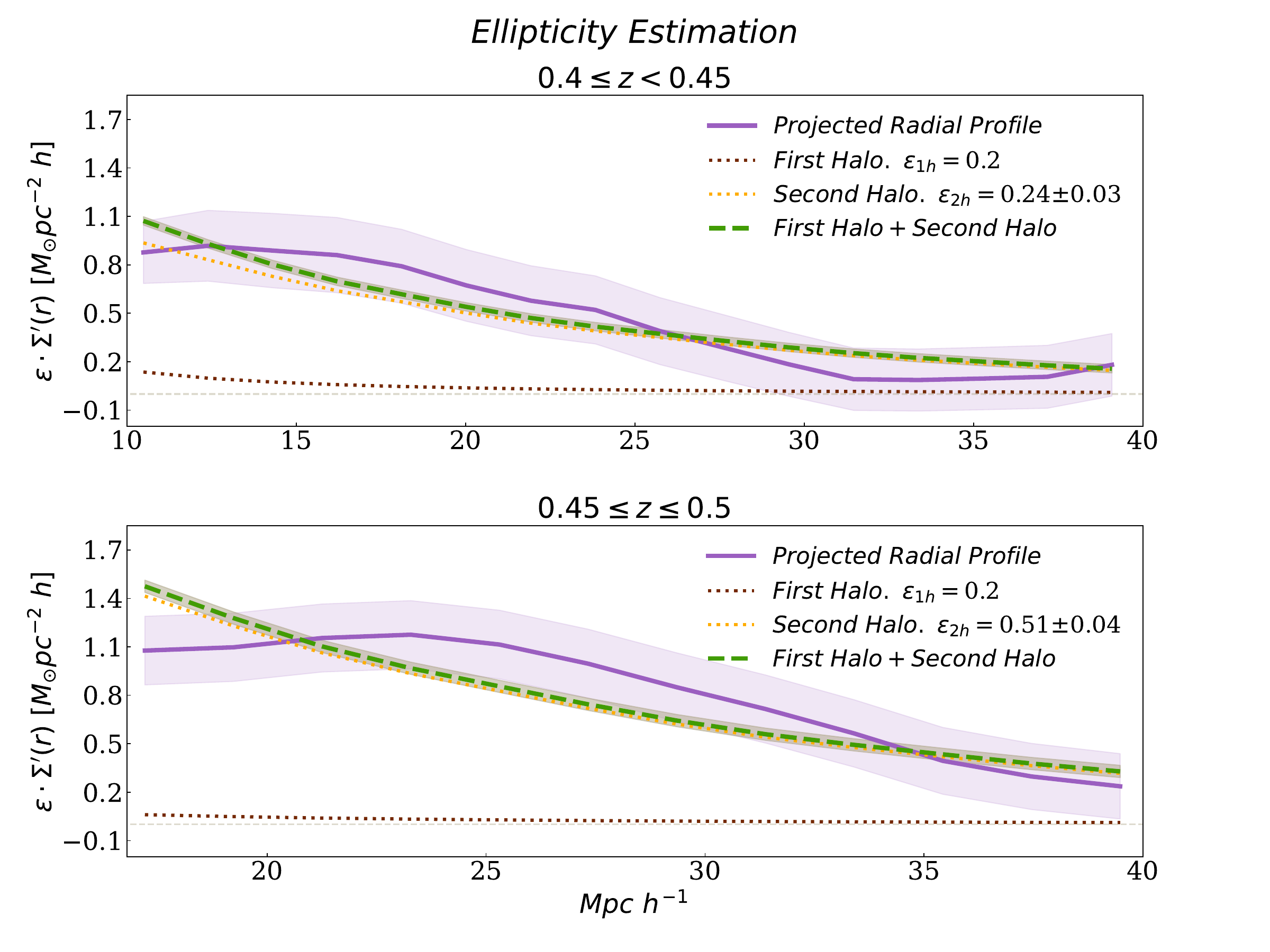}
    \caption{Fitting of the mean radial projected profile with respect to the anisotropic density quadrupolar model. In dashed green it can be seen the sum of the first halo term (dotted brown) and second halo term (dotted orange) of the anisotropic density model with ellipticities $\epsilon_{1h}$ and $\epsilon_{2h}$. In violet, the mean radial projected profile and its uncertainties calculated through Equation \ref{proyeccion_kappa}. For sample $S_1$ (upper panel), the ellipticity parameters are $\epsilon_{1h}=0.2$ and $\epsilon_{2h}= 0.24 \pm 0.03$ corresponding to a quasi-spherical halo shape. For sample $S_2$ (lower panel), the ellipticity parameters are $\epsilon_{1h}=0.2$ and $\epsilon_{2h}=0.51 \pm 0.04$ which is related to a elongated halo shape. The grey shaded region shows the fitting uncertainty.}
    \label{fig:z_vs_l2}
\end{figure}

In this section, we apply the anisotropic lensing model proposed by \cite{2022MNRAS.517.4827G} to model the obtained $\kappa$ convergence CMB data. 
The lensing effect produced by an ellipsoidal mass density can be studied through the surface mass density distribution $\Sigma (R)$ with $R$ the ellipsoidal radial coordinate $R^2=r^2(qcos^2(\theta)+sin^2(\theta)/q)$ with a semi-axis ratio $q \leq 1$ \citep{Uitert2017}. Following \cite{1996MNRAS.283..837S}, we define the ellipticity parameter $\epsilon= \frac{1-q}{1+q}$ in order to obtain a multipolar expansion:
\begin{gather} \label{model}
    \Sigma (R) \equiv \Sigma (r,\theta) := \Sigma_0(r) + \epsilon\Sigma_0'(r)cos(2\theta)
\end{gather}
where $\theta$ is the relative position angle of the source with respect to the major axis of the density mass distribution. $\Sigma_0$ y $\Sigma_0'$ are the monopolar and quadrupolar components, respectively. $\Sigma_0$ is related to the axis-symmetrical mass distribution while the quadrupole component is defined in terms of the monopole as:
\begin{equation} \label{quad}
    \Sigma_0' = -r \dfrac{d \ \Sigma_0(r)}{dr}
\end{equation}
corresponding to an elongated mass distribution. \\
Based on the halo model, the total surface density distribution can be described according to the sum of a first and second halo components.
The first halo term is modelled assuming a spherically symmetric NFW profile \citep{1997ApJ...490..493N}, which can be parameterized by the radius that encloses a mean density equal to 200 times the critical density of the Universe, $R_{200}$, and a dimensionless concentration parameter, $c_{200}$:
\begin{equation} \label{eq:nfw}
\rho_{1h}(r) =  \dfrac{\rho_{\rm crit} \delta_{c}}{(r/r_{s})(1+r/r_{s})^{2}},
\end{equation}
where  $r_{s}=R_{200}/c_{200}$ is the scale radius, $\rho_{\rm crit}$ is the critical density of the Universe and
$\delta_{c}$ is the characteristic overdensity described by:
\begin{equation}
\delta_{c} = \frac{200}{3} \dfrac{c_{200}^{3}}{\ln(1+c_{200})-c_{200}/(1+c_{200})}.  
\end{equation}
We define $M_{200}$ as the mass within $R_{200}$ that can be obtained as:
\begin{equation}
    M_{200}=200\,\rho_{\rm crit} (4/3)\pi\,R_{200}^{3}
\end{equation} 
To model this component we consider the mean mass specified in Table \ref{tab:samples_table} and we fix the concentration taking into account the relation given by \cite{2019ApJ...871..168D} which is in agreement with the \cite{2017MNRAS.466.3103S} model for the mass calculation.\\
The tridimensional density profile of the second halo term is modelled considering the halo-matter correlation function, $\xi_{hm}$, as:
\begin{equation} \label{eq:rho2h}
    \rho_{2h}(r) = \rho_{m} \xi_{hm} = \rho_{\rm crit} \Omega_m (1+z)^3 b(M_{200},\langle z \rangle) \xi_{mm}
\end{equation}
where $\rho_m$ is the mean density of the Universe, satisfying:
\begin{gather*}
    \rho_m = \rho_{\rm crit} \Omega_m (1+z)^3
\end{gather*}
and the halo-matter correlation function is related to the matter-matter correlation function through the halo bias \citep{2005PhRvD..71j3515S}:
\begin{gather*}
    \xi_{hm}  = b(M_{200},\langle z \rangle)\xi_{mm}
\end{gather*}
We set the halo bias by adopting \cite{2010ApJ...724..878T} model calibration.\\
Thus, the radial surface density distribution can be modeled taking into account essentially the monopolar and quadrupolar terms (Equation \ref{model}) of a main halo mass distribution, $\Sigma_{1h}$, and a contribution to the mass from the cluster neighborhood, $\Sigma_{2h}$, corresponding to a second halo term, with independent ellipticity parameters:
\begin{equation} \label{Sigma_R}
    \Sigma(R) \sim \Sigma_{1h}(r) + \epsilon_{1h}\Sigma_{1h}'(r)cos(2\theta)+\Sigma_{2h}(r)+\epsilon_{2h}\Sigma_{2h}'(r)cos(2\theta)
\end{equation}
This model assumes that the mass density distribution of the main halo as well as the second halo term are elongated along the same directions. Although a misalignment between the two halos terms can exist, this expected misalignment will bias the estimated elongation of the second halo term to lower values.\\
We derive estimators of cluster mass and ellipticity from the convergence maps following the procedures detailed in \cite{2022MNRAS.517.4827G} for the shear components. The model mass estimator arises from the integration of Equation \ref{Sigma_R} on $\theta$, which gives:
\begin{equation}
    \Sigma(r) = \Sigma_{1h}(r) + \Sigma_{2h}(r)
\end{equation}
This estimator can be compared to the observed radial $\kappa$ profile. \\
By multiplying Equation \ref{Sigma_R} by $cos(2\theta)$ and integrating in $\theta$, we obtain the model ellipticity estimators:
\begin{equation} \label{ellip}
    \Sigma_{proj}(r) = \epsilon_{1h}\Sigma_{1h}'(r) + \epsilon_{2h}\Sigma_{2h}'(r)
\end{equation}
which is related with the anisotropic contribution and can be compared with the observed lensing effect, by averaging the convergence $\kappa$ in angular bins:
\begin{gather} \label{proyeccion_kappa}
    \kappa_{proj} (r) = \frac{\sum_{j} \kappa_j (r) cos(2\theta_j)}{\sum_{j} cos^2(2\theta_j)}
\end{gather}
\\
In Figure \ref{fig:z_vs_l1} we show the adopted model for the monopole (Equation \ref{ellip}) together with the observed radial convergence profiles and the corresponding uncertainties. It can be seen the good behavior of the model in the central range of distances $[8, 40]$ Mpc$h^{-1}$ for the estimated mass derived from the richness-mass relation as a function of redshift for the redMaPPer clusters. It is worth to notice that our mass estimation through \cite{2017MNRAS.466.3103S} model, is in agreement with that given in \cite{geach2017} (for details, see their Figure 3) in the richness range for samples $S_1$ and $S_2$. \\
The low level agreement observed at small separations can be explained  from the gaussian smoothing of the map and the low number of pixels in the first bins. Also, it is important to take into account that in these regions there exists a SZ effect which decreases the signal and may affect our measurements. Besides, we notice that this model aims at the analysis of weak lensing by clusters for both, observations \citep{2021MNRAS.501.5239G} and simulations \citep{2022MNRAS.517.4827G}. Thus, several CMB lensing effects are not considered by this approach. Nevertheless, we acknowledge that the results are in a general good agreement in spite of the larger scales involved in our analysis.\\
We study the behavior of the quadrupolar terms, which are derived from the monopolar ones (Equation \ref{quad}), and their respective ellipticities. With this aim, we fit the projected radial profile (Equation \ref{proyeccion_kappa}) with respect to the ellipticity estimators (Equation \ref{ellip}). Given that we neglect in our analysis the internal region which is sensitive to the main halo component, we fix the ellipticity for this term at $\epsilon_{1h}=0.2$ which corresponds to a quasi-spherical halo shape and is in concordance with the previous redMaPPer ellipticity measurements \citep{2018MNRAS.475.2421S,2021MNRAS.501.5239G}. Taking this into account, we use non-linear least squares to fit the second halo term which has the largest contribution to the surface density profile in the analysed radial range. We estimate the uncertainty in $\epsilon_{2h}$ by means of the fitting covariance matrix. We notice that the estimated amplitude uncertainties are small,  within the measurement errors, providing suitable associated ellipticity values.
In Figure \ref{fig:z_vs_l2} we show the resulting mean radial profile for the theoretical quadrupolar terms $\epsilon_{1h}\Sigma_{1h}'(r)$ and $\epsilon_{2h}\Sigma_{2h}'$ with ellipticities $\epsilon_{1h}$ and $\epsilon_{2h}$ for samples $S_1$ and $S_2$. As it can be seen in this figure, samples $S_1$ and $S_2$ have different second halo ellipticity parameter values being:
\begin{gather}
\begin{split}
        \epsilon_{2h}= 0.24 \pm 0.03 \  , \ Sample \ S_1 \\
        \epsilon_{2h}=0.51 \pm 0.04 \ , \ Sample \ S_2
\end{split}
\end{gather}
Given the different redshift depth of samples $S_1$ and $S_2$, we fix the corresponding fitting regions $[10, 40]$, $[17, 40]$ Mpc$h^{-1}$, respectively. We also notice that in the high redshift sample, the fit is less accurate than for sample $S_1$, requiring a higher value of $\epsilon$. We argue that these effects are caused by the strong alignment of the high redshift clusters at scales where there is still a significant convergence signal in the parallel direction.

\section{Conclusions} \label{Conclusions}
Due to its large angular extension and depth, there is a great potential in the investigations of CMB lensing map for the analysis of the mass distribution at large scales. 
These studies provide important alternative information to weak gravitational lensing research, involving shear estimations by background galaxies.
In our work we perform a statistical study of the convergence $\kappa$ obtained from the temperature map of the Cosmic Microwave Background from the Planck 18 survey centered on rich galaxy clusters samples extracted from the redMaPPer catalogue. 
We find a significant excess in the convergence $\kappa$ values around clusters, which extends several tens of Mpc$h^{-1}$. We study this $\kappa$ profile for different richness samples as well as redshift intervals. We obtain, as expected, a larger amplitude for the higher richness samples and similar results for the redshift ranges explored. 
When clusters are aligned along their galaxy member distribution major axis, we find a significant difference between the $\kappa$ values along the parallel and perpendicular directions. Remarkably, this difference persists at scales as large as $40$ Mpc$h^{-1}$ indicating that this alignment is strongly conditioned by the large--scale mass anisotropic distribution. 
We study an anisotropic surface mass density model associated to the clusters which is in reasonable agreement with the observational results obtained. From this analysis we estimate the mean ellipticity of the second halo term for samples $S_1$, $\epsilon_{2h}= 0.24 \pm 0.03$ and $S_2$, $\epsilon_{2h}= 0.51 \pm 0.04$, consistent with an extended, highly flattened mass distribution oriented along the cluster major axis.

\section*{Acknowledgements}

This work was partially supported by the Consejo Nacional de Investigaciones Científicas y Técnicas (CONICET), the Agencia Nacional de Promoción Científica y Tecnológica (PICT-2020-SERIEA-01404) and the Secretaría de Ciencia y Tecnología (SeCyT), Universidad Nacional de Córdoba, Argentina. Also, it was supported by Ministerio de Ciencia e Innovación, Agencia Estatal de Investigación and FEDER (PID2021-123012NA-C44). This work is based on observations obtained with Planck \url{http://www.esa.int/Planck}), an ESA science mission with instruments and contributions directly funded by ESA Member States, NASA, and Canada. Some of the results in this paper have been derived using the HEALPix (K.M. Górski et al., 2005, ApJ, 622, p759) package. Plots were made using Python software and post-processed with Inkscape.  

\section*{Data Availability}

The data underlying this article will be shared on reasonable request. The data sets used in this article were derived from sources in the public domain. Data products and observations from Planck, \url{https://pla.esac.esa.int}.



\bibliographystyle{mnras}
\bibliography{example} 






\bsp	
\label{lastpage}
\end{document}